\documentclass[superscriptaddress, aps, preprintnumbers, twocolumn, amsmath, amssymb, sort&compress, nofootinbib, 10pt, paper, floatfix]{revtex4-2}

\usepackage{amsfonts,amsmath,amssymb,ascmac,bm,tensor, microtype}
\usepackage{fnpct} 
\usepackage{comment}
\usepackage{ifpdf}
\usepackage{slashed}
\usepackage{color}
\usepackage[mathscr]{eucal}
\usepackage[utf8]{inputenc}
\usepackage{cancel}
\usepackage{soul}
 
\ifpdf
  \usepackage{graphicx}     
  \usepackage[bookmarksopen,colorlinks=true,linkcolor=bblue,citecolor=bblue,urlcolor=ppink]{hyperref}
\else     
\fi

\definecolor{red}{rgb}{1,0,0}
\definecolor{darkred}{rgb}{0.6,0,0}
\definecolor{darkgreen}{rgb}{0.992447,0.623778,0.034597}
\definecolor{ppink}{rgb}{1,0.4,0.4}
\definecolor{bblue}{rgb}{0.284602,0.317763,0.963947}
\definecolor{purple}{rgb}{0.5 ,0, 0.7}

\definecolor{dgreen}{rgb}{0 ,0.5, 0.5}

\newcommand{\vev}[1]{ \left< {#1} \right> }
\newcommand{\dd}{\mathrm{d}}

\newcommand{\ee}{\text{e}}

\newcommand{\Pl}{{\text{Pl}}}

\newcommand{\dec}{{\text{dec}}}

\newcommand{\Mpc}{\text{Mpc}}

\newcommand{\Cyan}[1]{{\color{cyan}{#1}}}

\makeatletter
\newcommand\footnoteref[1]{\protected@xdef\@thefnmark{\ref{#1}}\@footnotemark}
\makeatother

\allowdisplaybreaks[1]

\begin{document}

\title{
Constraints on the Sharpness of the Curvature Power Spectrum
}

\author{Keisuke Inomata}
\affiliation{William H. Miller III Department of Physics and Astronomy, Johns Hopkins University, 3400 N. Charles Street, Baltimore, Maryland, 21218, USA}
\author{Xuheng Luo}
\affiliation{William H. Miller III Department of Physics and Astronomy, Johns Hopkins University, 3400 N. Charles Street, Baltimore, Maryland, 21218, USA}

\begin{abstract}
\noindent
Motivated by the fact that a sharply peaked curvature spectrum is often considered in the literature, we examine theoretical constraints on the sharpness of such a spectrum. In particular, we show that the sharply peaked curvature power spectrum, originating from the enhancement of subhorizon perturbations during inflation, is significantly constrained by energy conservation. While the constraints do not depend on the exact form of inflaton potential, we also study concrete inflaton potentials that realize a sharply peaked curvature spectrum and how theoretical limits are saturated in these cases.

\end{abstract}

\maketitle

\section{Introduction}

The curvature perturbation not only seeds all the structures we observe in the universe today, but also sheds light on its origins. On large scales ($ k \lesssim \mathcal O(1)\,\Mpc^{-1}$), the amplitude of the curvature perturbation, described by the curvature power spectrum $\mathcal P_\zeta$, has been precisely measured. Observations of the cosmic microwave background (CMB) anisotropy and the large-scale structure (LSS) of the universe indicate that $\mathcal P_\zeta \sim \mathcal O(10^{-9})$, with a slight scale dependence \cite{Planck:2018vyg}. However, on smaller scales ($ k \gtrsim \mathcal O(1)\,\Mpc^{-1}$), determining the curvature spectrum becomes increasingly challenging due to physical limitations such as Silk damping and nonlinear dynamics. Despite these challenges, numerous efforts have been made to measure or constrain the small-scale curvature power spectrum through various methods, including the UV galaxy luminosity function \cite{Sabti:2021unj}, strong gravitational lensing \cite{Gilman:2021gkj}, CMB distortions \cite{Chluba:2012we,Kohri:2014lza}, PBH production \cite{Josan:2009qn,Sato-Polito:2019hws,Yoo:2018kvb}, dynamical heating \cite{Graham:2023unf,Graham:2024hah}, scalar-induced GWs \cite{Inomata:2018epa,Dandoy:2023jot}, etc.

Since the curvature power spectrum is still observationally allowed to be large on small scales, this could lead to many detectable signals. For instance, large-amplitude curvature perturbations on small scales can produce a sizable amount of primordial black holes (PBHs)~\cite{Zeldovich:1967lct,Hawking:1971ei,Carr:1974nx,Carr:1975qj}, which may explain dark matter and/or the BHs detected by the LIGO-Virgo-KAGRA collaborations \cite{Bird:2016dcv,Sasaki:2016jop,Carr:2016drx,Kawasaki:2016pql,Inomata:2016rbd,Inomata:2017okj,Sasaki:2018dmp,Escriva:2022duf,Pritchard:2024vix}. 
In addition, these large curvature perturbations can induce substantial gravitational waves (GWs) through nonlinear interactions between tensor and scalar perturbations~\cite{10.1143/PTP.37.831,Matarrese:1993zf,Matarrese:1997ay,Ananda:2006af,Baumann:2007zm,Saito:2008jc,Saito:2009jt,Espinosa:2018eve,Kohri:2018awv,Inomata:2018epa,Domenech:2021ztg}, potentially explaining the recent detection of stochastic gravitational wave background (SGWB) by pulsar timing array (PTA) experiments \cite{NANOGrav:2023gor, NANOGrav:2023hde, NANOGrav:2023hvm,Antoniadis:2023rey, Antoniadis:2023utw, Antoniadis:2023zhi,Reardon:2023gzh, Zic:2023gta, Reardon:2023zen, Xu:2023wog}.

When studying the curvature perturbation on small scales, ansatzes such as Gaussian profiles and delta functions are commonly used to model the scale dependence of the curvature power spectrum \cite{NANOGrav:2023hvm,Kohri:2014lza,Inomata:2016rbd,Chluba:2012we,Yoo:2018kvb}. Although inflation models do not produce exactly the same curvature power spectrum as these ansatzes, it is generally believed that the physical observables do not strongly depend on the detailed shape of the power spectrum. Therefore, these ansatzes still serve as a good approximation for studying the underlying phenomena of the curvature power spectrum. However, a known exception is the induced GWs from a sharply peaked spectrum, where the extension of the infrared tail of the induced GW spectrum is sensitive to the sharpness of the curvature spectrum \cite{Pi:2020otn}. Importantly, the infrared tail could explain the SGWB recently detected by PTA experiments~\cite{NANOGrav:2023gor, NANOGrav:2023hde, NANOGrav:2023hvm,Antoniadis:2023rey, Antoniadis:2023utw, Antoniadis:2023zhi,Reardon:2023gzh, Zic:2023gta, Reardon:2023zen, Xu:2023wog}.
In particular, NANOGrav favored the power-law SGWB $ \Omega_{\text{GW}} \propto k^{1.8}$ \cite{NANOGrav:2023gor}, which is close to the shape of this infrared tail $\Omega_{\text{GW}} \propto k^2$ \cite{Kohri:2018awv,Pi:2020otn}\footnote{The similar slope can also be realized by other early universe phenomena \cite{Domenech:2024rks,Domenech:2019quo,Domenech:2020kqm,Harigaya:2023pmw,Bao:2024bws,Cui:2018rwi}.}. 
Given this, it is worth discussing the feasibility of such a sharply peaked spectrum appearing in inflation models.

In this paper, we examine theoretical constraints on a sharply peaked curvature power spectrum in single-field inflation models. We consider curvature power spectrum with a peaked feature on small scales. The peak of the power spectrum can be characterized by its amplitude $\mathcal P_\zeta (k_p)$, peak scale $k_p$, and width $\Delta k$. We put constraints on sharpness of the curvature power spectrum by considering both causality and energy conservation. Similarly to the uncertainty principle, a sharper power spectrum corresponds to a longer correlation of the two-point function in real space. However, due to causality, the real-space correlation cannot be larger than the Hubble horizon at the time when the peak rises. This limits the rise of the peak to very subhorizon scales if it is extremely sharp. Furthermore, enhancing perturbations on subhorizon scales requires a substantial amount of energy, which must be constrained by energy conservation~\cite{Inomata:2021zel}. The combination of these limitations puts tight constraints on the sharpness of the curvature spectrum, regardless of the specific form of the inflaton potential.

To see how the theoretical constraints work in concrete models, we present examples of inflaton potential that produce a sharply peaked curvature spectrum. In particular, we introduce a local oscillatory feature in the inflaton potential~\cite{Chen:2008wn}, which leads to resonant amplification of field perturbations within a very narrow range of wavelengths~\cite{Zhou:2020kkf,Peng:2021zon}. Such an oscillatory potential can, in principle, arise in theories where the inflaton field is also a pseudo-Goldstone boson \cite{Graham:2015cka,McAllister:2008hb,Flauger:2009ab}, or be effectively induced through interactions with oscillating spectator fields \cite{Inomata:2022ydj,Cui:2023fbg}. To further enhance such a sharp peak feature and generate observable signals within the theoretical constraints, we also modify the slope of the inflaton potential between the rise of the peak and the horizon exit of the peak.

The remainder of this paper is structured as follows. In Sec.~\ref{sec:2}, we discuss the theoretical constraints on the narrowly enhanced curvature power spectrum, and we show concrete potentials that realize such power spectrum in Sec.~\ref{sec:3}. In Sec.~\ref{sec:4}, we calculate the induced GW spectrum from the enhanced power spectrum, and we conclude in Sec.~\ref{sec:5}. Throughout this work, we take the following notation of the Fourier expansion for an arbitrary quantity $X$:
\begin{align}
    \label{eq:fourier}
  X(\bm x) = \int \frac{\dd^3 k}{(2\pi)^3} X_{\bm k} \ee^{i \bm k \cdot \bm x}. 
\end{align}

\section{ Lower bound on power spectrum sharpness}\label{sec:2}

\begin{figure}
        \centering \includegraphics[width=0.99\columnwidth]{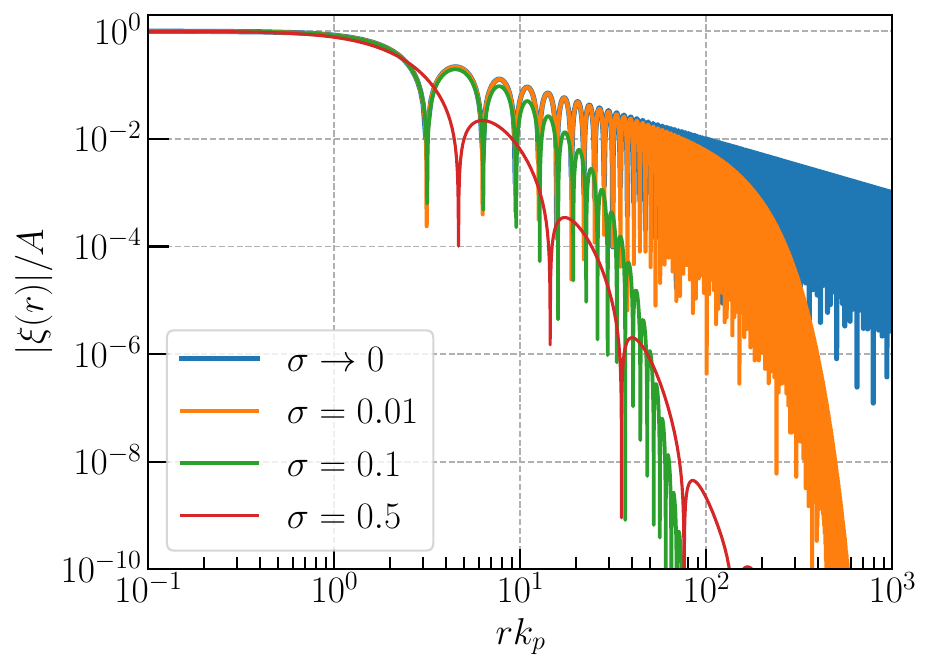}
        \caption{ 
        The correlation function defined by Eq.~(\ref{eq:xi}) with the power spectrum given by Eq.~(\ref{eq:gaussian}).
    }
        \label{fig:xi}
\end{figure}

In this section, we derive the theoretical lower bound on the sharpness of the power spectrum in single-field inflation without assuming a concrete form of the potential.
To this end, let us begin with the correlation function in real space:
\begin{align}
  \xi_X(r) \equiv \vev{X(\bm x) X(\bm x + \bm r)} = \int \dd \ln k \frac{\sin(kr)}{kr} \mathcal P_X(k),
  \label{eq:xi}
\end{align}
where the power spectrum is given by
\begin{align}
  \vev{X_{\bm k} X_{\bm k'}} = (2\pi)^3 \delta(\bm k + \bm k') \frac{2\pi^2}{k^3}\mathcal P_X(k).
\end{align}
To catch the essence, let us parameterize the sharply peaked power spectrum as 
\begin{align}\label{eq:gaussian}
  \mathcal P_X(k) = \frac{A}{\sqrt{2\pi} \sigma} \exp\left[- \left(\frac{\ln(k/k_p)}{\sqrt{2}\sigma}\right)^2 \right],
\end{align}
where the power spectrum is localized around $k_p$ with the width $\Delta k \sim k_p \sigma$.
Figure~\ref{fig:xi} shows the correlation function with different $\sigma$. 
From this figure, we can see that the real-space correlation decays around $r_\text{dec} \sim 1/(k_p\sigma)$. 
This can be understood as the uncertainty principle, $\Delta x \Delta k \sim 1$, where $\Delta k \sim k_p\sigma$ and $\Delta x \sim r_\text{dec}$ in our case. 
This behavior about the wavenumber width ($k_p\sigma$) and the decay radius ($r_\text{dec}$) must remain qualitatively the same even if we consider other shapes of the sharp power spectrum, such as the top-hat power spectrum.

Throughout this work, we discuss a sharply peaked power spectrum of curvature perturbations that arises from the enhancement of subhorizon perturbations $\delta \phi$ of the inflaton field $\phi$.
The rise of the sharp power spectrum must satisfy causality, which means that $\xi_{\delta \phi}(r)$ on the superhorizon scales must be suppressed during the rise. 
We denote the end of the rise of the sharp power spectrum by subscript ``$*$'', where the comoving Hubble scale is the largest during the rise.
The end of the rise can be the moment when the $\delta \phi$ enhancement stops or when the $\delta \phi$ enhancement is not narrowly limited around $k_p$. From causality, we can get the following inequality between the sharpness of the power spectrum $\sigma$ and the horizon scale $\mathcal H_*$:
\begin{align}
  r_\dec \sim \frac{1}{k_p\sigma} \lesssim \mathcal H_*^{-1},
  \label{eq:r_dec}
\end{align}
This indicates that, to realize the sharp power spectrum, its rise must occur on scales much smaller than the horizon scales \footnote{A peaked curvature power spectrum can be realized by transitions between slow-roll and ultra-slow-roll inflation without excitation of subhorizon modes \cite{Byrnes:2018txb,Inomata:2021tpx,Ng:2021hll}. However, bounded by causality, the width of such spectrum is not narrow ($\sigma \sim \mathcal O(1)$).}. Indeed, when we are considering actual inflation models in Sec.~\ref{sec:3}, Eq.~\eqref{eq:r_dec} is always implicitly satisfied. An example where Eq.~\eqref{eq:r_dec} is saturated up to $\mathcal O(1)$ factors is presented in Sec.~\ref{subsec:saturation}.

In addition to causality, we need to take into account energy conservation. 
The rise of a sharply peaked power spectrum must be realized through the energy transfer from the inflaton background to the enhanced perturbations because of energy conservation.
Then, the energy density of the perturbations at the end of the rise is given by~\cite{Inomata:2021zel}\footnote{Strictly speaking, the right-hand-side of Eq.~(\ref{eq:rho_dphi}) should include the potential term, $V'' \delta \phi^2_*$ ($V'' \equiv \dd^2 V/\dd \phi^2$). Here we assume that around the $\phi_*$, there is at least one point where the potential term can be safely ignored.}
\begin{align}\label{eq:rho_dphi}
  \rho_{\delta \phi} &\approx \frac{1}{2a^2_*} \vev{a_*^2 (\delta \dot\phi_*)^2 + (\partial_i \delta \phi_*)^2} \nonumber \\
  &\sim \frac{k^2_p}{a^2_*} \vev{\delta \phi_*^2} \nonumber \\
  &\sim  \frac{k^2_p}{a^2_*} \sigma \mathcal P_{\delta \phi}(k_p,\eta_*),
\end{align}
where the dot means the physical time derivative and we have used in the last line that the variance of the perturbations in real space is proportional to the integral of curvature power spectrum over $\ln k$ and hence $\vev{\delta \phi_*^2}  \sim \sigma \mathcal P_{\delta \phi}(k_p,\eta_*)$. Let us denote the start of the rise by subscript $0$. From energy conservation, $\rho_{\delta \phi} $ must be smaller than the decrease of the total background energy of inflaton during the rise, which yields
\begin{align}\label{eq:rhodelta}
  \rho_{\delta \phi} < \left( \frac{\dot{\phi}^2_0}{2} +V(\phi_0)\right) - \left(\frac{\dot{\phi}^2_*}{2} +V(\phi_*) \right).
\end{align}
We consider the situation where \footnote{Suppose the rise occurs on the Hubble timescale, Eq.~\eqref{eq:deltaV} implies that the averaged $\epsilon$ during the rise ($\Bar{\epsilon}$) is nearly equal to or less than $\epsilon_0$. The violation of Eq.~(\ref{eq:deltaV}) corresponds to a large decrease in potential energy during the rise.
In this case, the $\epsilon_0$ in Eq.~\eqref{eq:rho_const} should be replaced by the averaged $\Bar{\epsilon}$ during the rise to account for the large potential decrease.
}
\begin{equation}\label{eq:deltaV}
     V(\phi_0) - V(\phi_*) \lesssim \epsilon_0 H^2 M_\Pl^2
\end{equation}
here $\epsilon \equiv - \dot H/H^2$, and at the start of the rise, $2\epsilon_0H^2M_\Pl^2 \approx \dot{\phi}^2_0 $. The energy conservation in Eq.~\eqref{eq:rhodelta} implies that 
\begin{align}
  \rho_{\delta \phi} \lesssim \epsilon_0 H^2 M_\Pl^2.
  \label{eq:rho_const}
\end{align}
 Combining Eqs.~(\ref{eq:rho_dphi}) and (\ref{eq:rho_const}), we obtain 
\begin{align}
  \frac{k^2_p}{a^2_*}\sigma \mathcal P_{\delta \phi}(k_p,\eta_*) \lesssim \epsilon_0 H^2 M_\Pl^2.
  \label{eq:p_const}
\end{align}

The curvature perturbation is related to $\mathcal P_{\delta \phi}(k_p,\eta_*)$ as
\begin{align}
  \mathcal P_\zeta(k_p) \approx \frac{1}{2\epsilon_{\text{exit}} M_\Pl^2} Q \left( \frac{a_*}{a_{\text{exit}}}\right)^2 \mathcal P_{\delta \phi}(k_p,\eta_*),
\end{align}
where the subscript ``exit" denotes the value at the horizon exit of the peak scale $k_p$. Note that the factor $ Q (a_*/a_{\text{exit}})^2$ takes into account the evolution of $\mathcal P_{\delta \phi}(k_p)$ from the rise until its horizon exit, where $Q = 1$ if the slow-roll inflation holds after the rise. We will explain below Eq.~(\ref{eq:el_e0}) when $Q>1$ is realized.
Substituting this into Eq.~(\ref{eq:p_const}), we obtain 
\begin{align}
   &  2   \frac{k^2_p}{\mathcal H_*^2} Q^{-1}\left(\frac{a_{\text{exit}}}{a_*}\right)^2 \frac{\epsilon_{\text{exit}}}{\epsilon_0} \sigma\mathcal P_{\zeta}(k_p) \lesssim 1.\nonumber \\
  & \Rightarrow \   \left( \frac{k_p}{\mathcal H_*}\right)^4 Q^{-1}\frac{\epsilon_{\text{exit}}}{\epsilon_0} \sigma\mathcal P_{\zeta}(k_p) \lesssim 1,
  \label{eq:pzeta_bound}
\end{align}
 where in the second line, we have used $a_{\text{exit}}/a_* \approx k_p/\mathcal H_*$ to relate the rise and the horizon exit, and we have neglected the $\mathcal O(1)$ factor.
Using Eq.~(\ref{eq:r_dec}), we finally obtain 
\begin{align}\label{eq:sharpness}
  &\sigma^{-3} Q^{-1}\frac{\epsilon_{\text{exit}}}{\epsilon_0} \mathcal P_{\zeta}(k_p) \lesssim 1 \nonumber \\
  \Rightarrow \ & 
  \sigma \gtrsim \left(Q^{-1}\frac{\epsilon_{\text{exit}}}{\epsilon_0} \sigma \mathcal P_{\zeta}(k_p)\right)^{1/4}.
\end{align}
We keep $\sigma \mathcal P_\zeta$ in the right-hand-side because the physical quantities in real space, such as the two point function, are determined by the integral over $\ln k$ and hence $\sigma \mathcal P_\zeta$ is more physical than $\mathcal P_\zeta$ itself. For the parametrization in Eq.~\eqref{eq:gaussian}, $A \sim \sigma \mathcal P_\zeta(k_p)$. Eq.~(\ref{eq:sharpness}) is the constraint on the sharpness of the curvature power spectrum in single-field inflation models.

\section{Concrete examples}\label{sec:3}

In this section, we discuss the inflation models that realize the sharp curvature power spectrum, satisfying the condition in Eq.~(\ref{eq:sharpness}).
Specifically, we consider a model where the inflaton potential, $V(\phi) \equiv V_{\text{b}}(\phi)+V_{\text{osc}}(\phi)$, consists of a local oscillatory feature $V_{\text{osc}}(\phi)$, superimposed on a smooth base potential $V_{\text{b}}(\phi)$. The equation of motion of $\phi$ is given by
\begin{equation}\label{eq:bg_eom}
    \Ddot{\phi} + 3H \Dot{\phi} + V'(\phi) = 0,
\end{equation}
where $V'(\phi) = \dd V(\phi)/\dd\phi$.
Before encountering the oscillatory potential, the field $\phi$ slow-rolls down the smooth base potential $V_{\text{b}}(\phi)$ following
\begin{equation}\label{eq:mean}
    \dot{\phi} \approx -\frac{V'_{\text{b}}(\phi)}{3H} \approx \sqrt{2\epsilon}M_\Pl H,
\end{equation}
where we have assumed $V'_\text{b} < 0$ without loss of generality. As the inflaton field enters the oscillatory regime of the potential, the mean value of the field, $\phi_{\text{mean}}$\Cyan{,} continues to evolve following Eq.~\eqref{eq:mean}, but with small and rapid oscillations $\phi_{\text{osc}}$ superimposed due to $V_{\text{osc}}(\phi)$.

We further require that $\Dot{\phi}_{\text{mean}} \gg |\Dot{\phi}_{\text{osc}} | $ in the oscillatory regime,  ensuring that the evolution of $\phi$ can be approximated as
\begin{equation}\label{eq:phi_approx}
    \phi(t)  \approx \sqrt{2\epsilon}M_\Pl H (t- t_0) +\phi(t_0).
\end{equation}
One can estimate the magnitude of $|\Dot{\phi}_{\text{osc}} |$ based on Eq.~\eqref{eq:bg_eom}. Suppose that the characteristic oscillatory timescale of $\phi_{\text{osc}}$ is $\Lambda/H$ with $\Lambda \ll 1$, then $\Ddot{\phi}_{\text{osc}} \sim H\Dot{\phi}_{\text{osc}}/\Lambda  \gg  H \Dot{\phi}_{\text{osc}}$, and
\begin{equation}\label{eq:small_osc}
    V_{\text{osc}}'(\phi) \approx -\ddot{\phi}_{\text{osc}}\sim H\Dot{\phi}_{\text{osc}}/\Lambda.
\end{equation}
Comparing with Eq.~\eqref{eq:mean}, we find that $\Dot{\phi}_{\text{mean}} \gg |\Dot{\phi}_{\text{osc}} | $ if 
\begin{equation}\label{eq:linearphi}
    \left| \frac{V_{\text{osc}}'(\phi)}{V_{\text{b}}'(\phi)} \right| \ll \Lambda^{-1}.
\end{equation}
For the remainder of this section, we require that Eq.~\eqref{eq:linearphi} is satisfied, which corresponds to $c\ll 1$ in the parameterization used in Eqs.~\eqref{eq:potential1} and \eqref{eq:potential2}.

Before encountering the oscillatory potential, the fluctuations of $\phi$ follow the Bunch-Davies vacuum condition. Upon entering the oscillatory regime, the fluctuations of $\phi$ acquire an effective mass equal to $\dd^2V_{\text{osc}}/\dd\phi^2$, which also oscillates over time. If the physical wavelength of the fluctuation $\sim a/k$ matches the oscillation timescale of the effective mass $\sim \Lambda /H$, resonant amplification can occur in fluctuations \cite{Inomata:2022yte}. The growth of the subhorizon fluctuations can be described by the following linear equation of motion
\begin{equation}\label{eq:phi_pt_eom}
\delta \ddot{\phi}_{\bm k}+3 H \delta \dot{\phi}_{\bm k}+\left(\frac{k^2}{a^2}+V''\right) \delta \phi_{\bm k}=0,
\end{equation}
where $V'' = \dd^2 V(\phi)/\dd \phi^2$, and $\phi$ can be calculated from Eq.~\eqref{eq:bg_eom}.
The criterion for resonant amplification and its dependence on different $k$ vary with the choice of potential $V_{\text{osc}}(\phi)$. Therefore, to continue our discussion, we study two representative inflaton potentials that exhibit resonant amplification.

\begin{figure*}
    \centering
    \includegraphics[width=0.49\linewidth]{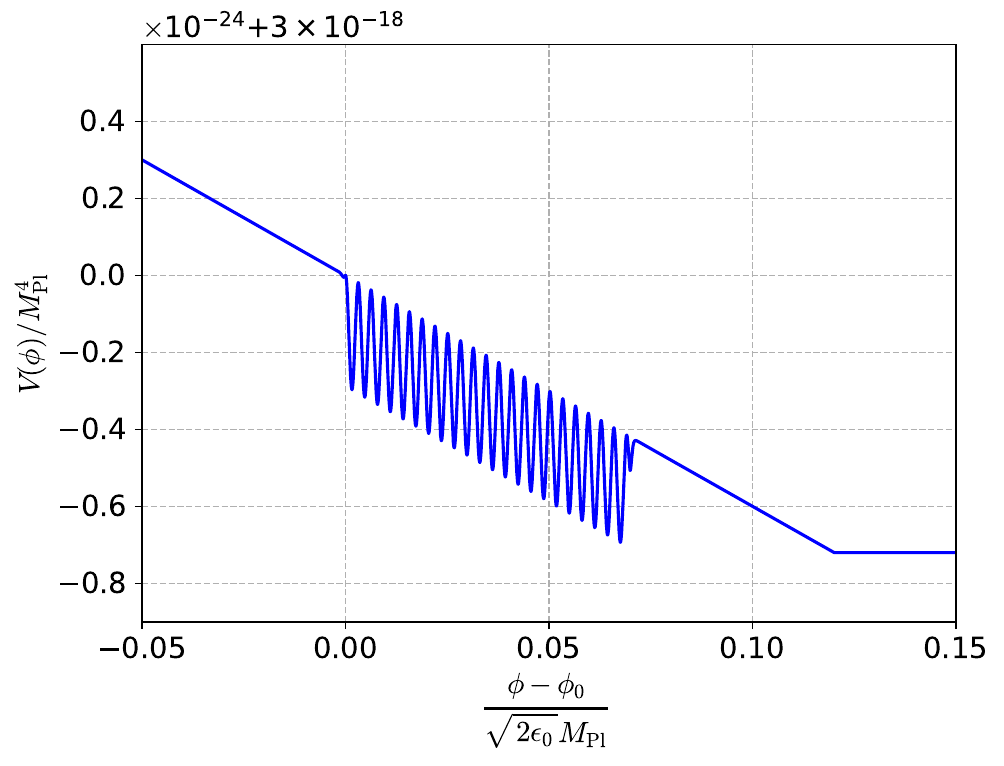}
    \includegraphics[width=0.49\linewidth]{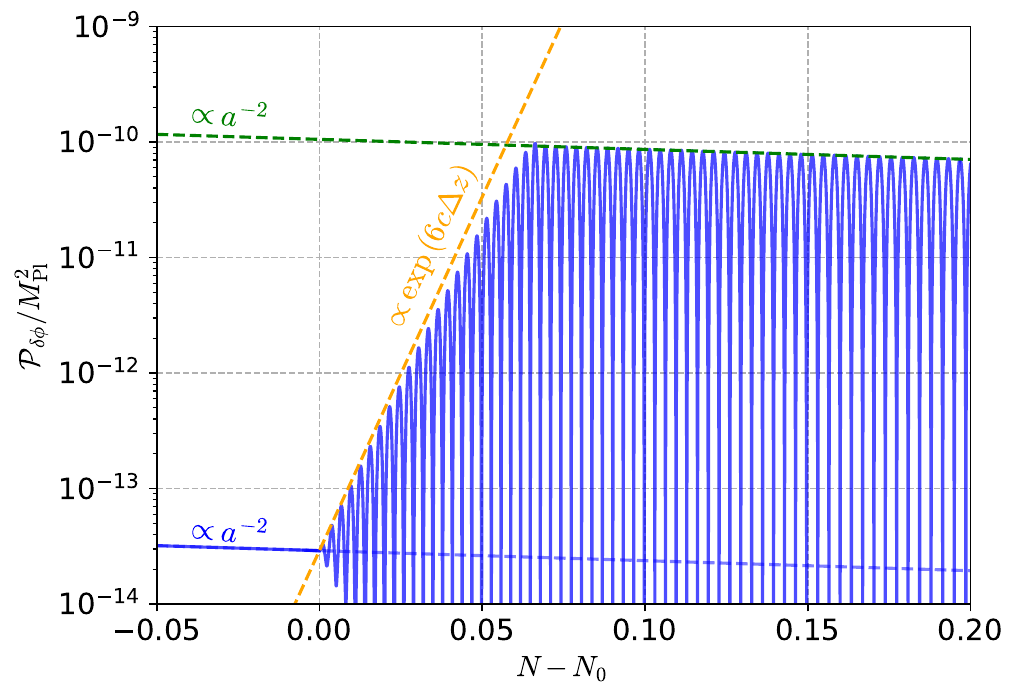}\\
   \includegraphics[width=0.49\linewidth]{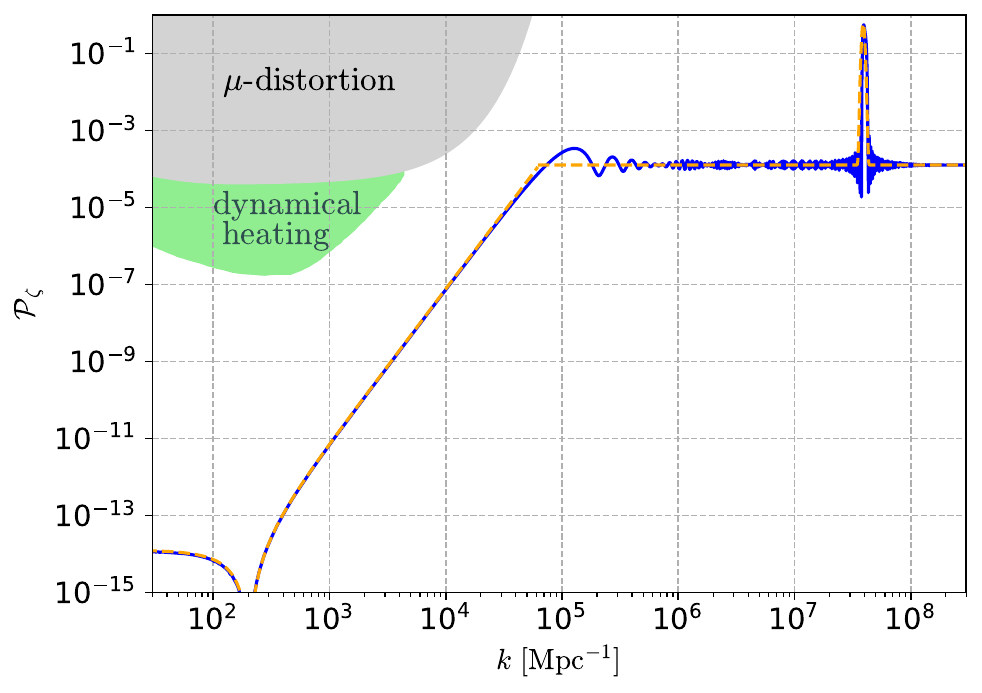} \includegraphics[width=0.49\linewidth]{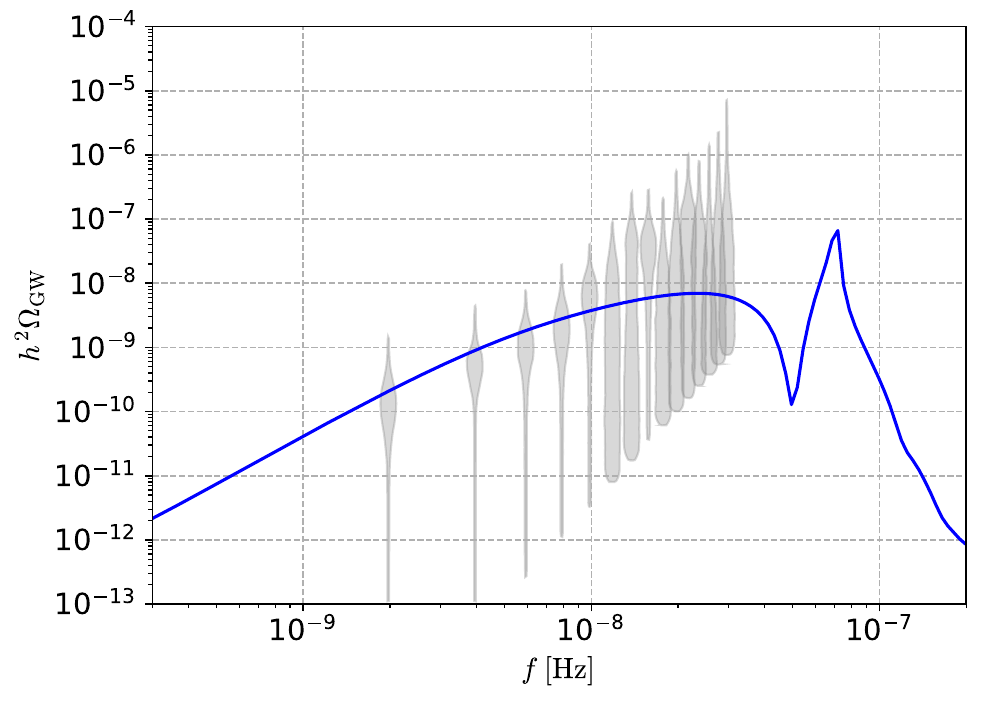}
    \caption{The results from case 1 with parameters (Eq.~\eqref{eq:potential1}): $c = 0.024$, $\Lambda = 5 \times 10^{-4}$, $N_{\text{osc}} = 0.07$, $V_0/M_\Pl^4 = 3\times 10^{-18}$, $\epsilon_0 = 10^{-6}$, $\epsilon_l = 10^{-16}$, $\frac{\phi_l - \phi_*}{\sqrt{2\epsilon_0}M_\Pl}= 0.05$, and $\mathcal H _0 = 3.64 \times 10^4 \text{ Mpc}^{-1}$. The parameters are chosen such that the peak part of the power spectrum has the width and height that is within the $68\%$ Bayesian credible region of the Gaussian peak model reported in Ref.~\cite{NANOGrav:2023hvm}, while saturating the theoretical bound in Eq.~\eqref{eq:sharpness2} up to $\mathcal O(1)$ factors. \textit{Top Left: } The potential form given by Eq.~\eqref{eq:potential1}. \textit{Top Right: }  The resonant amplification of $\mathcal P_{\delta \phi}$ at the peak scale $k_p = 3.85 \times 10^7 \text{ Mpc}^{-1}$ obtained by numerically solving the equation of motion of $\delta\phi$, Eq.~(\ref{eq:phi_pt_eom}). The horizontal axis is the e-folds from the time when $\phi = \phi_0$. The analytical approximation of the growth is shown by the orange dashed line, based on Eq.~\eqref{eq:ana_growth}. \textit{Bottom Left: } The curvature power spectrum. The blue line represents the numerical result and the orange dashed line represents the analytical result from Eqs.~\eqref{eq:cps} and \eqref{eq:ana_cps}, where we take $k_{\text{dip}}=2.01\times 10^2 \text{ Mpc}^{-1} \approx 1.55(\epsilon_l/\epsilon_0)^{1/4} \mathcal H_l$  and $\kappa = 0.24$. The shaded regimes illustrate existing bounds from COBE/FIRAS via $\mu$-distortions \cite{Inomata:2016rbd,Chluba:2012we,Kohri:2014lza} and from dynamical heating of ultrafaint dwarf galaxies \cite{Graham:2023unf,Graham:2024hah}. \textit{Bottom Right: } The spectrum of induced GWs with the curvature power spectrum given in the bottom left panel. The GW spectrum is calculated with Eq.~\eqref{eq:omegaGW}. The gray violin plots correspond to the reconstructed GW background spectrum from NANOGrav \cite{NANOGrav:2023hvm}.}
    \label{fig:2}
\end{figure*}

\subsection{Case 1: An oscillatory feature with a fixed period}\label{subsec:case1}
In the first case, we consider an inflaton potential that has a local oscillatory feature with fixed period in field value:
\begin{align}\label{eq:potential1}
  V(\phi) &= V_b(\phi) + 2V_0 c\epsilon_0 \, D(\phi) \left(-1 + \cos\left( \frac{\phi-\phi_0}{\sqrt{2\epsilon_0} \Lambda M_\Pl} \right) \right)\nonumber \\
  &\quad + V_{\text{end}}(\phi),
\end{align}
where $D$ is the smooth top-hat function for $\phi_0 \lesssim \phi \lesssim \phi_*$, defined as
\begin{align}
  D(\phi) &\equiv \left(\frac{1 + \tanh\left(\frac{\phi-\phi_0}{\sqrt{2\epsilon_0} \Lambda M_\Pl}\right)}{2}\right) \left(\frac{1 + \tanh\left(\frac{\phi_* - \phi}{\sqrt{2\epsilon_0} \Lambda M_\Pl}\right)}{2} \right).
\end{align}
$V_\text{end}$ is the modification of the potential around the end of inflation, which is irrelevant to the resonant amplification, and therefore we neglect it in the following.
The base potential is given by
\begin{align}
  V_b(\phi) = V_0\left[1 - \sqrt{2\epsilon(\phi)} \frac{\phi-\phi_0}{M_\Pl} \right].
\end{align}
We include the possible change of $\epsilon$ between the rise of the peak and the horizon exit of the peak in
\begin{align}
  \epsilon(\phi) = \epsilon_0 + \left(\frac{1 + \tanh\left(\frac{\phi - \phi_l}{\sqrt{2\epsilon_0} \Lambda M_\Pl}\right)}{2} \right) (\epsilon_l - \epsilon_0),
  \label{eq:el_e0}
\end{align}
where we introduced $\phi_l$ and $\epsilon_l$ to characterize the change of $\epsilon$, and the slope of the potential changes from $\epsilon_0$ to $\epsilon_l \ll \epsilon_0$ at $\phi_l > \phi_*$. During evolution, $\epsilon \approx \epsilon_0$ before $\phi \sim \phi_l$, after which $\epsilon$ gradually decreases to $\epsilon_l$ at the rate $\propto a^{-6}$. The top left panel of Fig.~\ref{fig:2} shows the potential form with the fiducial parameter sets described in its caption.
In general, a sharply peaked curvature spectrum that leads to observable induced GW or PBH signals requires the further enhancement of the curvature perturbation achieved by decreasing $\epsilon$ or increasing $Q$. This is evident in Eq.~\eqref{eq:sharpness}. For example, if we consider $\sigma \mathcal P_\zeta(k_p) \gtrsim 10^{-4}$ with $Q = 1$ and $\epsilon$ fixed, Eq.~(\ref{eq:sharpness}) leads to $\sigma \gtrsim 0.1$. While a large $Q$ can be achieved by introducing a broader parametric resonance after the very narrow resonance, in this section, we only consider decreasing $\epsilon$ and leaving exploration of the another possibility for future studies. Given the lack of significant enhancement of the perturbation after the rise of the peak, $Q = 1$ holds for the remainder of this section.

In the absence of the oscillatory potential ($c = 0$), the curvature power spectrum produced by the potential in Eq.~\eqref{eq:potential1} is approximately given by \cite{Zegeye:2021yml}
\begin{equation}\label{eq:cps}
\mathcal P_{\zeta,\text{base}}(k) \approx 
\begin{cases}
\frac{H^2}{8\pi^2M_\Pl^2\epsilon_l} & \ k \gtrsim \mathcal H_l
\\[2mm]
\frac{H^2}{8\pi^2M_\Pl^2\epsilon_0}\left( 1-(k/k_{\text{dip}})^2 \right)^2 & \ k \lesssim \mathcal H_l
\end{cases},
\end{equation}
where $\mathcal H_l$ is the conformal Hubble parameter at $\phi = \phi_l$ and $k_{\text{dip}} \sim (\epsilon_l/\epsilon_0)^{1/4} \mathcal H_l$. We require $\mathcal H_l \lesssim k_p$ so that the sharp peak of the curvature spectrum is further enhanced by the decrease of $\epsilon$.

 Now we switch to discussing the evolution of $\delta \phi$ based on the analysis in Ref.~\cite{Inomata:2022yte}. For simplicity, we define $N_{\text{osc}} \equiv \frac{\phi_* - \phi_0}{\sqrt{2\epsilon_0}M_\Pl}$, which is approximately the number of e-folds for $\phi$ to roll over the oscillatory regime from $\phi_0$ to $\phi_*$. Suppose $\phi(t_0) = \phi_0$, Eq.~\eqref{eq:phi_pt_eom} can be rewritten into a Mathieu-like equation
\begin{align}\label{eq:mathieu}
    \frac{\mathrm{d}^2 \Phi_{\bm k}(z)}{\mathrm{d} z^2}+\left[\alpha_k(z)-2 q \cos (2 z)\right] \Phi_{\bm k}(z)=0, 
\end{align}
where we have ignored higher order terms, including those related to $D(\phi)$, and defined
\begin{align}\label{eq:ana_define}
    \Phi_{\bm k}\equiv a^{3/2} \delta \phi_{\bm k},\  2 z \equiv \frac{H}{\Lambda}(t-t_0), \nonumber \\ \alpha_k(z)=4 \Lambda^2\left[\left(\frac{k}{aH} \right)^2 -\frac{9}{4}\right],\  q=6 c .
\end{align}
Exponential growth of $\Phi_{\bm k}$ occurs if
\begin{align}\label{eq:sharp1}
    1-q\lesssim \alpha_k(z) \lesssim 1+q.
\end{align}
The perturbation growth is most significant at wavenumber $k \approx \mathcal H/2\Lambda$ with growth rate characterized by the Floquet index $\mu_k \approx q/2$ to be 
\begin{align}\label{eq:ana_growth}
    \Phi_{\bm k} &\propto \exp{(\mu_k (z_* - z_0))} \propto \exp{(3c\Delta z)},
\end{align}
where $\Delta z \equiv  z_*- z_0 \approx N_{\text{osc}}/2\Lambda$ with $z_0$ and $z_*$ being the values of $z$ at $\phi = \phi_0$ and $\phi_*$, respectively (we similarly denote $a_0$ and $\mathcal H_0$ in this way in the following).
The top right panel of Fig.~\ref{fig:2} shows the evolution of $\delta \phi_{\bm k_p}$ with the fiducial parameter, where note that $\mathcal P_{\delta \phi} \propto \delta \phi^2$.

The realization of a sharp power spectrum requires significant enhancement of $\delta \phi$ over a narrow wavenumber $\Delta k \sim \sigma k_p$. This places a limit on the choice of parameters $c,\Lambda$ and $N_{\text{osc}}$. First, from Eq.~\eqref{eq:sharp1}, one observes that 
\begin{equation}\label{eq:sharp2}
    c \lesssim \sigma, \ N_{\text{osc}} \lesssim \sigma
\end{equation}
are required. This is because: first, increasing $c$ results in a broadening of the range of $k$ that satisfies Eq.~\eqref{eq:sharp1}; second, since the wavenumber at the center of the enhancement ($k \approx aH/2\Lambda$) increases with $a$, increasing $N_{\text{osc}}$ will also broaden the range of $k$ that is enhanced. Furthermore, from Eq.~\eqref{eq:ana_growth}, a significant enhancement of the perturbation requires $c\Delta z \gtrsim 1$. Combined with Eq.~\eqref{eq:sharp2} and $\Delta z  \approx N_{\text{osc}}/2\Lambda$, this condition leads to the upper bound on $\Lambda$:
\begin{equation}\label{eq:sharp3}
    \Lambda \lesssim \sigma^2.
\end{equation}

Using Eqs.~\eqref{eq:sharp1} and \eqref{eq:ana_growth}, we find that the resultant sharply peaked curvature power spectrum can be analytically approximated by the form in Eq.~\eqref{eq:gaussian} with 
\begin{equation}\label{eq:ana_cps}
    \frac{\Delta \mathcal  P_{\zeta}(k)}{\mathcal P_{\zeta,\text{base}}(k) } \approx \kappa e^{6c\Delta z- 3N_{\text{osc}}} \exp\left[- \left(\frac{\ln(k/k_p)}{\sqrt{2}\sigma}\right)^2 \right],
\end{equation}
where $k_p \approx \mathcal H_0(1+N_{\text{osc}})/2\Lambda$ (this is slightly more accurate than setting $k_p \approx \mathcal H_0/2\Lambda$), $\sigma \approx c$, $\Delta \mathcal  P_{\zeta}(k) \equiv \mathcal P_{\zeta}(k) - \mathcal P_{\zeta,\text{base}}(k)$  and $\mathcal P_{\zeta,\text{base}}(k)$ is given in Eq.~\eqref{eq:cps}. The amplification at the peak is estimated based on the maximum Floquet index $\mu_k \approx q/2$. This results in an overestimation of the peak that we absorb in the constant $\kappa \sim   \mathcal O (0.1)$. Eq.~\eqref{eq:ana_cps} combined with Eq.~\eqref{eq:cps} yields our analytical prediction of the curvature power spectrum generated from Eq.~\eqref{eq:potential1}, which is a sharp peak on top of a (roughly) step-like curvature spectrum. We plot the numerically derived $\mathcal P_{\zeta}(k)$ by solving Eq.~\eqref{eq:phi_pt_eom} together with the analytical expression of $\mathcal P_{\zeta}(k)$ in the bottom left panel of Fig.~\ref{fig:2}.

As Eq.~\eqref{eq:sharpness} is derived independently of specific inflaton potentials, a stronger limit on $\sigma$ can apply when considering an exact model. For the concrete case we have discussed (case 1), Eq.~\eqref{eq:sharp3} sets the bound on the wavenumber that can realize a sharp power spectrum with width $\sigma$ as $k_p\approx \mathcal H_0/2\Lambda \gtrsim  \mathcal H_0/\sigma^2$, which appears when Eqs.~\eqref{eq:sharp2} and \eqref{eq:sharp3} are saturated. This relation is a more restrictive bound on $k_p$ than the causality constraint in Eq.~\eqref{eq:r_dec}. This means that excitation of more subhorizon modes than that in Eq.~\eqref{eq:r_dec} is required to achieve the sharp spectrum, which should be even more constrained by the energy conservation relation in Eq.~\eqref{eq:pzeta_bound}. Substituting $k_p \gtrsim  \mathcal H_0/\sigma^2$ into Eq.~\eqref{eq:pzeta_bound} yields\footnote{In Ref.~\cite{Inomata:2022yte}, the authors discussed higher-order corrections to the linear perturbation theory in a model with a large amplification of field perturbations during inflation. 
As discussed in it, the energy conservation constraints discussed here are of the same order as the constraints for the subdominant loop power spectrum discussed in Ref.~\cite{Inomata:2022yte}. Since we discuss the orders by neglecting the $\mathcal O(1)$ factors throughout this work, we focus only on the energy conservation bound in our analysis.
See also Ref.~\cite{Caravano:2024tlp} for the lattice analysis on fully nonperturbative cases.
}
\begin{align}\label{eq:sharpness2}
  \sigma \gtrsim \left(\frac{\epsilon_l}{\epsilon_0} \sigma \mathcal P_{\zeta}(k_p)\right)^{1/8},
\end{align}
where we have used $\mathcal H_* \approx \mathcal H_0$ and $\epsilon_* \approx \epsilon_0$ because with $\Delta N_{\text{osc}} \ll 1$, the beginning and end of resonant amplification are almost simultaneous. From Eqs.~\eqref{eq:cps} and \eqref{eq:ana_cps}, we find $\mathcal P_{\zeta}(k_p) > \mathcal P_{\zeta,\text{base}}(k \gg \mathcal H_l) \approx \epsilon_0 \mathcal P_{\zeta,\text{base}}(k \ll \mathcal H_l)/\epsilon_l $. By substituting this into Eq.~\eqref{eq:sharpness2}, we obtain
\begin{align}\label{eq:case1_sharpness}
    \sigma > \mathcal P_{\zeta,\text{base}}(k \ll \mathcal H_l)^{1/7}.
\end{align}
 Suppose that the same potential also generates the curvature power spectrum observed by CMB measurement: $\mathcal P_{\zeta,\text{base}}(k \ll \mathcal H_l) \sim \mathcal P_{\zeta}(k_{\text{CMB}} = 0.05 \text{ Mpc}^{-1}) \approx 2.1\times 10^{-9}$ \cite{Planck:2018jri}. Eq.~\eqref{eq:case1_sharpness} demands that to satisfy energy conservation, the sharply peaked spectrum generated from Eq.~\eqref{eq:potential1} must have $\sigma > 0.1$, which is consistent with what we observed when numerically scanning the model parameter space in Eq.~\eqref{eq:potential1}. Since the curvature power spectrum is only measured at very large scales, it is possible to relax the limitation in Eq.~\eqref{eq:case1_sharpness} by suppressing the intermediate curvature power spectrum at $k_{\text{CMB}} \ll k \ll \mathcal H_l$. This can be achieved ad hoc by significantly increasing $\epsilon$ between the epoch that generates the CMB fluctuations and the epoch that generates the sharp peak spectrum. Note that we here only aim to find a proof-of-existence case that generates a sharply peaked spectrum. To avoid further complication of our analysis, 
 we only consider in Fig.~\ref{fig:2} the curvature spectrum at wavelength much smaller than $k_{\text{CMB}}^{-1}$ and assume that the CMB power spectrum is already generated by the earlier part of the inflation.

\begin{figure*}
    \centering
    \includegraphics[width=0.49\linewidth]{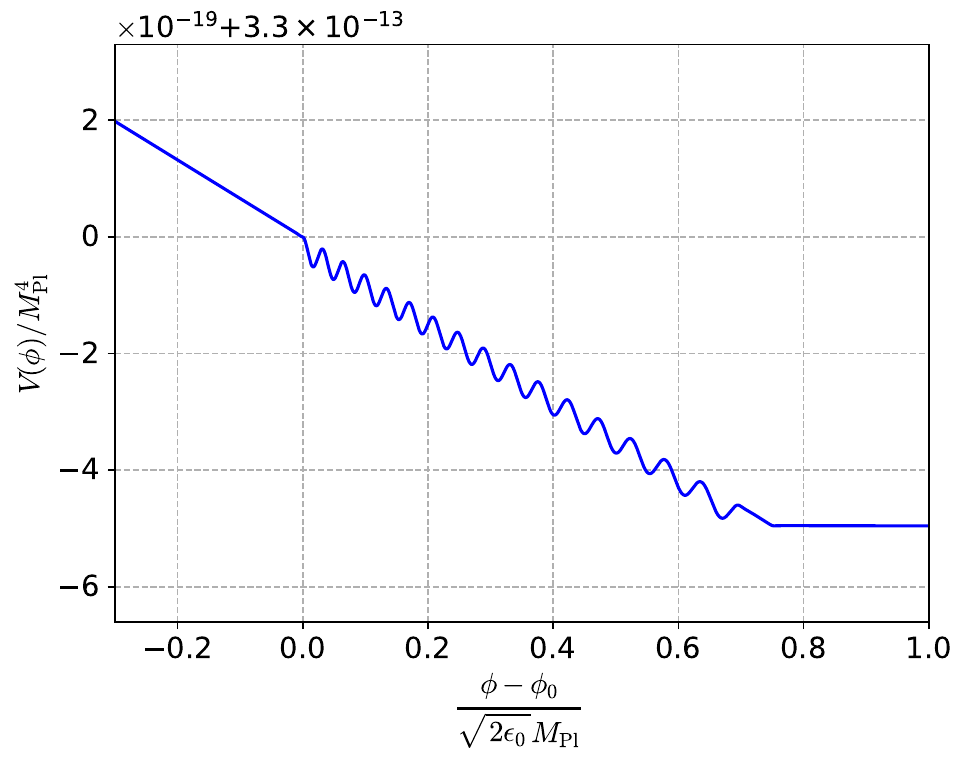}
    \includegraphics[width=0.49\linewidth]{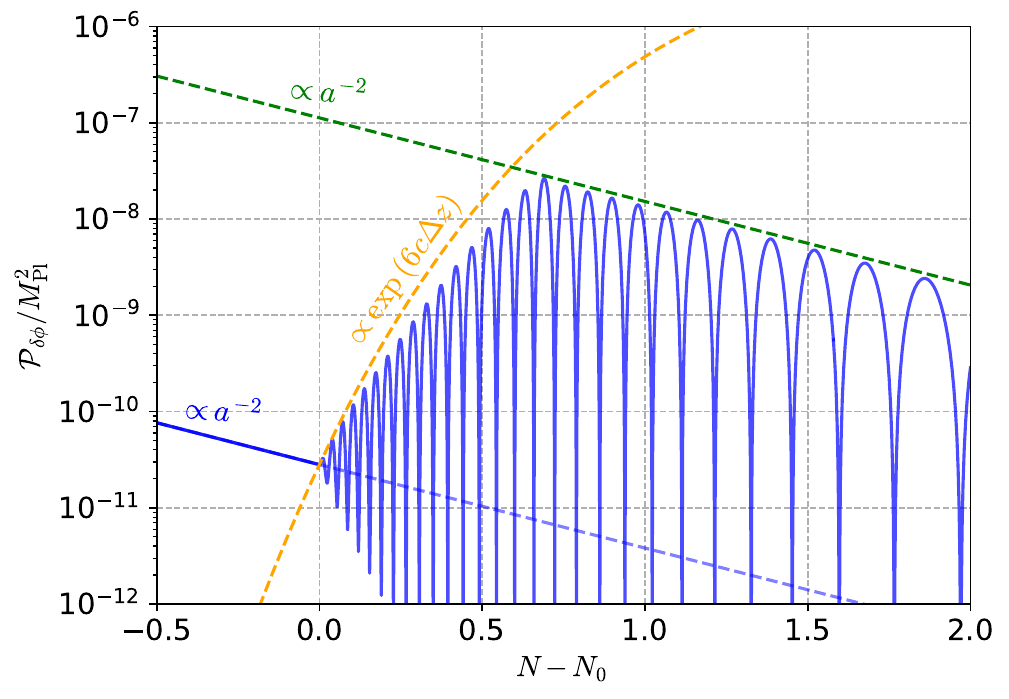}\\
   \includegraphics[width=0.49\linewidth]{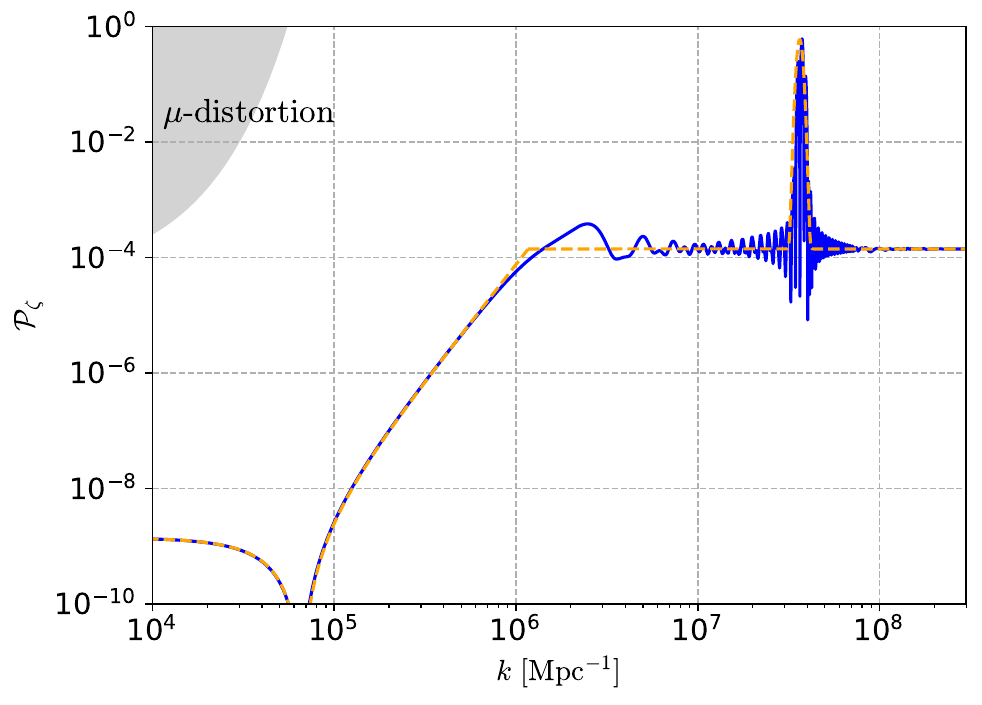} \includegraphics[width=0.49\linewidth]{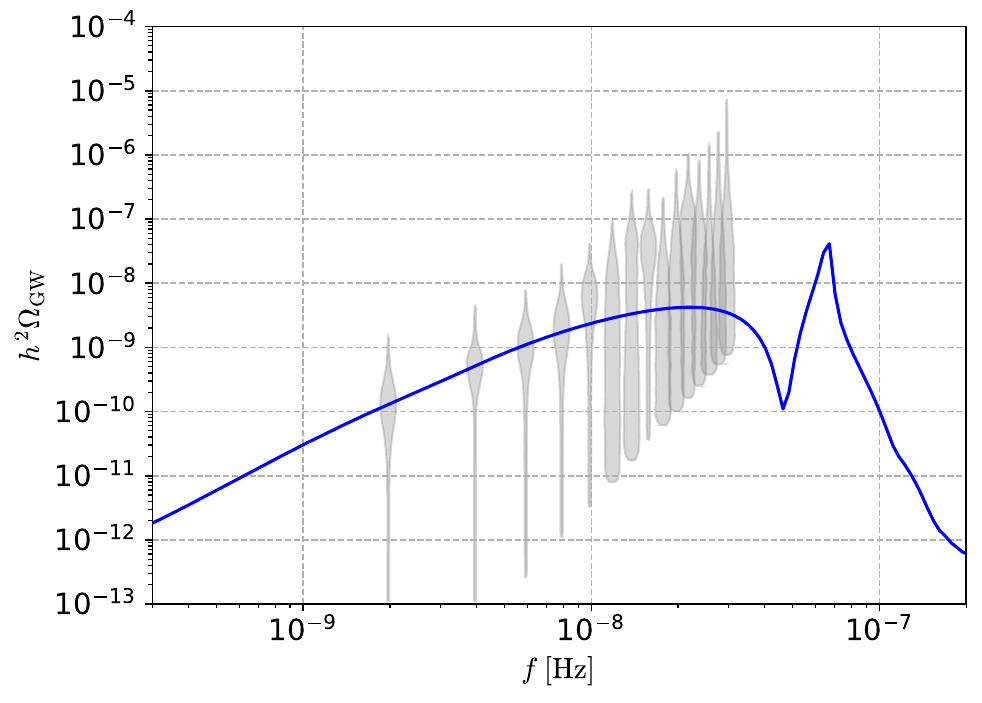}
    \caption{Similar to Fig.~\ref{fig:2} but for case 2 (Eq.~\eqref{eq:potential2}) with parameters: $c = 0.031$, $\Lambda = 5 \times 10^{-3}$, $N_{\text{osc}} = 0.7$, $V_0/M_\Pl^4 = 3.3\times 10^{-13}$, $\epsilon_0 = 10^{-6}$, $\epsilon_l = 10^{-11}$, $\frac{\phi_l - \phi_*}{\sqrt{2\epsilon_0}M_\Pl}= 0.05$, and $\mathcal H _0 = 3.64 \times 10^5 \text{ Mpc}^{-1}$. Here both Eqs.~\eqref{eq:r_dec} and \eqref{eq:pzeta_bound} are saturated up to $\mathcal O(1)$ factors. We take $k_p = 3.64 \times 10^7 \text{ Mpc}^{-1}$, $k_{\text{dip}}=6.55\times 10^4 \text{ Mpc}^{-1} \approx 1.51(\epsilon_l/\epsilon_0)^{1/4} \mathcal H_l$, and $\kappa = 1.7$.} 
    \label{fig:3}
\end{figure*}

\subsection{Case 2: An oscillatory feature with a gradual increase in period}\label{subsec:saturation}
In the previous case, the oscillatory feature of the potential occurs in a fixed period of $\phi$. As $\phi$ rolls over the oscillatory regime, $\dd^2V/\dd\phi^2$ oscillates over a period of proper time, eventually resulting in resonant amplification at the corresponding physical wavelength. We demonstrated previously that the on-resonance comoving wavelength will gradually redshift out of the on-resonance regime due to cosmic expansion. As a result, the duration of the resonance is limited to $N_{\text{osc}} \lesssim \sigma$ in Eq.~\eqref{eq:sharp2}, which eventually becomes a restrictive bound on the peak scale $k_p$. Since the model discussed in the previous section is unable to saturate Eq.~(\ref{eq:r_dec}), for completion of our discussion, we explore the potential form where both the causality constraint in Eq.~(\ref{eq:r_dec}) and the energy conservation constraint in Eq.~(\ref{eq:pzeta_bound}) saturate at the same time:
\begin{align}\label{eq:potential2}
  V(\phi) &= 2V_0 c\epsilon_0 \, D(\phi) \left(-1 + \cos\left( \frac{1- \exp{\left(\frac{\phi_0-\phi}{\sqrt{2\epsilon_0} M_\Pl}\right)}}{ \Lambda} \right) \right)\nonumber \\
  &\quad + V_b(\phi) + V_{\text{end}}(\phi).
\end{align}
The above potential is defined such that it is exactly the same as in Eq.~\eqref{eq:potential1} when the exponential is expanded in linear order of $\phi_0-\phi$. With the exponential, the oscillation period of the potential gradually increases to compensate for the redshift of the on-resonance mode. This results in a resonant amplification that focuses on a comoving scale, rather than a physical scale as in case 1. To observe this, one can rewrite the equation of motion of $\delta \phi$ similarly to Eq.~\eqref{eq:mathieu}:
\begin{align}\label{eq:mathieu2}
    \frac{\mathrm{d}^2 \Phi_{\bm k}(z)}{\mathrm{d} z^2}+\left[\alpha_k(z)-2 q \cos (2 z)\right] \Phi_{\bm k}(z)=0, 
\end{align}
where the definition of $\Phi_{\bm k}$ and $z$ are changed to
\begin{align}\label{eq:ana_define2}
    \Phi_{\bm k} \equiv a \delta \phi_{\bm k},\  2 z \equiv \frac{1-e^{H(t_0-t)}}{\Lambda}, \nonumber \\ \alpha_k(z)=4 \Lambda^2\left[\left(\frac{k}{\mathcal H_0} \right)^2 -\frac{2a^2}{a^2_0}\right],\  q=6 c .
\end{align}
The on-resonance mode is given by $k \approx \mathcal H_0/2\Lambda$ similar to case 1, but now increasing $N_{\text{osc}}$ does not broaden the on-resonance regime. To simplify our analysis, we conservatively require $N_{\text{osc}} \lesssim 1$ such that the beginning and end of resonant amplification are within one Hubble time $a_0 \sim a_*$. Following the same analysis in Sec.~\ref{subsec:case1}, we find that the sharp power spectrum with $\Delta k/k_p = \sigma$ can be realized under weaker conditions compared to case 1:
\begin{align}
    c \lesssim \sigma, \ \Lambda \lesssim \sigma.
\end{align}
Therefore, for case 2, the wavenumber that can realize a sharp power
spectrum with width $\sigma$ is bounded as $k_p\approx  \mathcal H_0/2\Lambda \gtrsim  \mathcal H_0/\sigma$, which can saturate the relation in Eq.~\eqref{eq:r_dec} up to $\mathcal O(1)$ factors.

The curvature power spectrum generated in case 2 is very similar to case 1: a step-like curvature spectrum with a sharp peak at the top. Similar to Eq.~\eqref{eq:ana_cps}, the analytical expression for case 2 is
\begin{equation}\label{eq:ana_cps2}
    \frac{\Delta \mathcal  P_{\zeta}(k)}{\mathcal P_{\zeta,\text{base}}(k) } \approx \kappa e^{6c\Delta z- 2N_{\text{osc}}} \exp\left[- \left(\frac{\ln(k/k_p)}{\sqrt{2}\sigma}\right)^2 \right],
\end{equation}
where $\mathcal P_{\zeta,\text{base}}(k)$ is the same as case 1, given in Eq.~\eqref{eq:cps}. However, since the definition of $\Phi_k$ and $z$ has changed from Eq.~\eqref{eq:ana_define} to Eq.~\eqref{eq:ana_define2}, the following changes to Eq.~\eqref{eq:ana_cps2} are required: $\Delta z \approx (1-\exp{(-N_{\text{osc}})})/2\Lambda$, $k_p = \mathcal H_0 /2\Lambda$ and replace $e^{-3N_{\text{osc}}}$ in Eq.~\eqref{eq:ana_cps} with $e^{-2N_{\text{osc}}}$. The resultant enhanced curvature power spectrum from both numerical and analytical calculations is illustrated in Fig.~\ref{fig:3}. We note that the analytical predictions of the evolution of $\mathcal P_{\delta \phi}$ in the top right panel of Figs.~\ref{fig:2} and \ref{fig:3} are different due to the definition change of $z$.

\section{Induced gravitational wave spectrum}\label{sec:4}
In this section, we discuss the production of GWs from the sharply peaked curvature power spectrum. Tensor perturbations (SGWB) are induced by curvature (scalar) perturbations at second order in the perturbation theory~\cite{10.1143/PTP.37.831,Matarrese:1993zf,Matarrese:1997ay,Ananda:2006af,Baumann:2007zm,Saito:2008jc,Saito:2009jt,Espinosa:2018eve,Kohri:2018awv,Inomata:2018epa,Domenech:2021ztg}. This compelling source of SGWB can be probed by current and future GW observations (see Ref.~\cite{Schmitz:2020syl} for the summary of the current and future GW sensitivities of experiments).
Although GWs can in principle be induced both during and after inflation, those induced during inflation are suppressed by the slow-roll parameter $\epsilon$~\cite{Inomata:2021zel}.
Given this, we focus on the GWs induced after the inflation in the following.
In particular, we consider the case where the sharp power spectrum enters the horizon during radiation domination for simplicity\footnote{See Refs.~\cite{Inomata:2019zqy,Inomata:2019ivs} for the complexity of the GWs induced by the curvature perturbations that enter the horizon during matter domination.}.
Then, the GW production during radiation domination can be calculated by \cite{Kohri:2018awv} \footnote{In Ref.~\cite{Garcia-Saenz:2022tzu}, trispectrum-induced SGWB has been shown to be negligible compared to the one in Eq.~\eqref{eq:omegaGW} as long as the theory is perturbative. Since, throughout this work, we keep our prediction of curvature perturbation and field perturbation in the perturbative regime, we expect higher order contribution to the GW production such as trispectrum-induced-GW to be at most subleading.}
\begin{widetext}
    \begin{align}\label{eq:omegaGW}
        \Omega_{\text{GW,RD}}(k) \equiv \frac{\dd\rho_{\text{GW}}/\dd\ln{k}}{\rho_{\text{tot,RD}}} \approx \int_0^{\infty} \mathrm{d} t \int_{-1}^1 \mathrm{~d} s\   \overline{I_{\text{RD}}^2(t, s)} \mathcal{P}_\zeta(k(t+s+1)/2) \mathcal{P}_\zeta(k (t-s+1)/2),
    \end{align}
where
    \begin{align}
        \overline{I_{\text{RD}}^2(t, s)}= & \frac{24\left(t(2+t)(s^2-1)\right)^2\left(-5+s^2+t(2+t)\right)^2}{(1-s+t)^8(1+s+t)^8}\left(\frac{\pi^2}{4}\left(-5+s^2+t(2+t)\right)^2 \Theta(t-(\sqrt{3}-1))\right. \nonumber \\
        & \left.+\left(-(t-s+1)(t+s+1)+\frac{1}{2}\left(-5+s^2+t(2+t)\right) \log \left|\frac{-2+t(2+t)}{3-s^2}\right|\right)^2\right),
    \end{align}
\end{widetext}
here $\Theta$ is the Heaviside step function. The energy density of GW observed today is related to $\Omega_{\text{GW,RD}}(k)$ by \cite{Inomata:2023zup}
\begin{equation}
    h^2\Omega_{\text{GW}}(f) = 0.43 \left( \frac{g_{*,\rho}}{80} \right) \left( \frac{g_{*,s}}{80} \right)^{-4/3} h^2\Omega_{r,0}\Omega_{\text{GW,RD}}(k),
\end{equation}
where the GW frequency is related to the comoving wavenumber by $f = k/(2\pi)$, $g_{*,\rho}$ ($g_{*,s}$) denote the effective numbers of relativistic degrees of freedom for the energy (entropy) densities when modes with $k$ reenter the horizon, and the radiation energy density parameter is given by $h^2\Omega_{r,0} \approx 4.2\times 10^{-5}$ from Planck 2018 data release~\cite{Planck:2018vyg}.

It is known that the induced GW spectrum associated with the sharply peaked curvature power spectrum has a distinctive feature. Specifically, $\Omega_{\text{GW}}$ produced by a sharp Gaussian peak curvature spectrum in the long-wavelength regime is $\propto k^2$ for $k \gtrsim 2\sigma k_p$ and $\propto k^3$ for $k \lesssim 2\sigma k_p$~\cite{Pi:2020otn}.\footnote{The transition scale $k \sim \sigma k_p$ is associated with the correlation length of the scalar perturbations in real space. Recall the discussion below Eq.~(\ref{eq:gaussian}).}
In the short-wavelength regime, a characteristic peak arises at $k \approx 2k_p/\sqrt{3} $, and the spectrum cutoff is located around $k \approx 2k_p$~\cite{Saito:2008jc,Saito:2009jt}. We numerically calculate the GW spectrum using Eq.~\eqref{eq:omegaGW} with the curvature power spectrum obtained in the previous section, and the results are shown in the bottom right panels in Figs.~\ref{fig:2} and \ref{fig:3}. We find that the resultant GW spectrum indeed shows the qualitative feature discussed above. In addition to this, since the $\mathcal{P}_\zeta(k)$ also consists of a flat curvature power spectrum $\mathcal P_{\zeta,\text{base}}$, in Eq.~\eqref{eq:omegaGW}, $\mathcal{P}_\zeta \times \mathcal{P}_\zeta$ also includes a $\mathcal P_{\zeta,\text{base}} \times \mathcal P_{\zeta,\text{base}}$ term and a $\mathcal P_{\zeta,\text{base}} \times \mathcal P_{\zeta,\text{peak}}$ term. The contribution to the GW spectrum from the $\mathcal P_{\zeta,\text{base}} \times \mathcal P_{\zeta,\text{base}}$ term is approximately $\Omega_{\text{GW}} \sim  \Omega_{r,0} \mathcal P_{\zeta,\text{base}}^2$. For the parameter sets in Figs.~\ref{fig:2} and \ref{fig:3}, $\mathcal P_{\zeta,\text{base}} \lesssim 10^{-4}$, thereby contributing negligibly to $\Omega_{\text{GW}}$ in the wavelength regime presented there. On the other hand, the cross-term $\mathcal P_{\zeta,\text{base}} \times \mathcal P_{\zeta,\text{peak}}$ is dominant at $k \gtrsim 2\sigma k_p$ where the drop in the GW spectrum becomes more gradual than the case where $\mathcal P_{\zeta,\text{base}} $ is absent.

Let us here connect the induced GW spectrum to the stochastic GW background (SGWB) recently detected by the PTA observations: NANOGrav~\cite{NANOGrav:2023gor, NANOGrav:2023hde, NANOGrav:2023hvm},   EPTA/InPTA~\cite{Antoniadis:2023rey, Antoniadis:2023utw, Antoniadis:2023zhi}, PPTA~\cite{Reardon:2023gzh, Zic:2023gta, Reardon:2023zen}, and CPTA~\cite{Xu:2023wog}. In particular, NANOGrav favors a power-law SGWB with $ \Omega_{\text{GW}} \propto k^{1.8}$ \cite{NANOGrav:2023gor}, which is close to $\propto k^2$. As mentioned above, $k^2$ can be naturally realized by the long-wavelength tail of the GW spectrum induced by the sharply peaked curvature power spectrum \cite{NANOGrav:2023hvm,Inomata:2023zup}. We illustrate the SGWB spectrum obtained from our models in Figs.~\ref{fig:2} and \ref{fig:3} together with the reconstructed SGWB spectrum from NANOGrav in Ref.~\cite{NANOGrav:2023hvm}. The peak of the GW spectrum induced by a curvature power spectrum parametrized by Eq.~\eqref{eq:gaussian} is roughly speaking $\Omega_{\text{GW}}(k_p) \sim \mathcal P_{\zeta}^2(k_p)\sigma^2 \Omega_{r,0}$. As NANOGrav detects $\Omega_{\text{GW}} \sim 10^{-8}$ at the peak, $\mathcal P_{\zeta}(k_p)\sigma \gtrsim 0.01$ is required to generate enough SGWB.

PBHs can form after an enhanced curvature perturbation enters the Hubble horizon. Since an enhanced curvature power spectrum with $\mathcal P_{\zeta}(k_p)\sigma \gtrsim 0.01$ is required to explain the GW observed by NANOGrav, these power spectra can also produce a substantial amount of PBHs. However, the PBH production rate has a huge uncertainty depending on the calculation scheme \cite{Franciolini:2023pbf, Iovino:2024uxp}. 
As a sanity check, we internally calculate the abundance of PBHs using the approach in Ref.~\cite{Inomata:2023zup} which is similar to the method used in Ref.~\cite{NANOGrav:2023hvm} and do not find an overproduction of PBHs. 
Given that the main focus of this work lies in the theoretical constraints of the sharp power spectrum, a full discussion of PBH production is beyond the scope of this paper.

\section{Conclusion}\label{sec:5}

A sharply peaked curvature spectrum, although often assumed in the context of the induced GW, dark matter halo, and PBH, is under a couple of theoretical constraints, which we investigate in this work. In particular, we have examined the narrowly enhanced curvature spectrum that arises from a single-field inflation. We have pointed out that such a sharp peak feature is limited by causality to occur at subhorizon scales (Eq.~\eqref{eq:r_dec}). Moreover, the enhancement of perturbations at subhorizon scales is strongly constrained by energy conservation (Eq.~\eqref{eq:pzeta_bound}). The combination of these two physical constraints imposes a lower bound on the sharpness of the curvature power spectrum.

Then, we have discussed inflaton potentials that realize a sharply peaked curvature spectrum. In particular, we consider an inflaton potential that has a local oscillatory feature characterized by its amplitude $c$, period $\Lambda$, and width $N_{\text{osc}}$. Then, the requirements on the sharpness of the spectrum can be expressed as those on these three parameters. To be concrete, we have considered two cases. In the first case, we added a local oscillatory feature with a fixed period in the field value and found that the model can produce a sharp spectrum, but cannot saturate the causality bound in Eq.~(\ref{eq:r_dec}). To study the case where both the causality bound in Eq.~(\ref{eq:r_dec}) and the energy conservation bound in Eq.~(\ref{eq:pzeta_bound}) are saturated at the same time, we considered the second case in which the period of the oscillatory potential increases as the inflaton field evolves. Then, we indeed found that both bounds are saturated in this case. For both cases, we numerically calculated the evolution of the inflaton field and its perturbation with results consistent with analytical estimates. To illustrate our results, we picked a set of parameters for each case which result in a sharply peaked curvature spectrum. We then numerically calculated the GWs produced by these curvature spectra and found that they can be consistent with the SGWB observed by the PTA observations.

Throughout this paper, we have assumed single-field inflation. However, the results can be generalized as long as one field dominates the evolution of the universe during the relevant periods, specifically from the rise of the peak to its horizon exit. Conversely, we have not considered the scenarios with non-standard initial states, such as warm inflation \cite{Graham_2009, Berera:1995ie, Lyth:1995hj, Lyth:1995ka, Ji:2021mvg,Tanin:2017bzm,Berghaus:2019whh,Berghaus:2024zfg} or cases where higher derivative interactions are important \cite{Garriga:1999vw,Chen:2006nt,Holman:2007na}. Extending our analysis to more generalized scenarios is left for future studies.

\acknowledgments
\noindent
We thank Marc Kamionkowski for useful discussions.
K.\,I.\, was supported by JSPS Postdoctoral Fellowships for Research Abroad.

\small
\bibliographystyle{apsrev4-1}
\bibliography{sharp_ps.bib}

\end{document}